\documentstyle[12pt]{article}
\def\beq{\begin{equation}}
\def\brr{\begin{array}}
\def\err{\end{array}}
\def\eeq{\end{equation}}
\def\bea{\begin{eqnarray}}
\def\eea{\end{eqnarray}}
\def\bs{\bigskip}
\def\ni{\noindent}

\def\nn{\nonumber}
\def\ms{\medskip}
\def\sp{\mbox{Sp}}
\def\txs{\textstyle}

\begin{document}

\hfill UB-ECM-PF 92/6
\mbox{}

\vspace*{1cm}

\begin{center}

{\LARGE \bf
Spontaneous compactification in 2D induced quantum gravity}

\vspace{1cm}

{\sc E. Elizalde and S.D. Odintsov}\footnote{On leave from Department of
Mathematics and Physics,
 Pedagogical Institute, 634041 Tomsk, Russia.}
%\\ \mbox{} \\
\ms

Department E.C.M., Faculty of Physics, \\
University of Barcelona, \\
Diagonal 647, 08028 Barcelona, Spain \\
{\it e-mail: eli @ ebubecm1.bitnet}

\vspace{1cm}

{\sl March 1992}

\vspace{2cm}

{\bf Abstract}

\end{center}

Spontaneous compactification ---on a $R^1\times S^1$ background--- in 2D
induced quantum gravity (considered as a toy model for more fundamental
quantum gravity) is analyzed in the gauge-independent effective action
formalism. It is shown that such compactification is stable, in
contradistinction to multidimensional quantum gravity on a
$R^D\times S^1 \ (D>2)$ background ---which is known
 to be one-loop unstable.

%\noindent{PACS: 04.60.+n, 12.25.+e, 12.10.-g, 11.15.Bt}

\newpage

\section{Introduction}
\ms

The  exact solution of 2D induced quantum gravity in the
light-cone gauge
or in the conformal gauge [1] has originated a number of works dealing
with
this theory [2-4] (and further references therein). Different
approaches, based on conformal field theory [1], on random matrix
models [2], and on topological field theory [3], have been
developed.
Presently, we see that there is some agreement among the exact
results
obtained from these different approaches. Thus, we have an exactly
solvable 2D quantum gravity, which can be considered as a toy model
for
the much more complicated 4D quantum gravity ---where so far only the
perturbative viewpoint is quite developed. For this reason, it would
be useful to formulate the perturbative approach to 2D induced
gravity
[5-7]. Some interesting results on this line have been obtained. For
instance, the one-loop calculation of the counterterms in 2D
induced
gravity has been done in different gauges [5-7], and renormalizability
has been found for some models.
The covariant gauge, in which the counterterms disappear, has also
been
found [5].
Thus, developing the perturbative approach to 2D induced gravity,
one
can expect to find an indication (at least qualitative) of some new
phenomena which would also take place in 4D induced gravity, in the
well-known language of usual, perturbative field theory. In other words,
2D quantum gravity can provide us with important information on general
properties of quantum gravity (a good example are the quantum
corrections to 2D black holes [8]).

Some years ago, in the spirit of the Kaluza-Klein approach, the
quantum
spontaneous-compactification program (also called self-consistent
dimensional reduction) for multidimensional quantum gravity was
developed (for a comprehensive introduction and general review see
[9]). The gauge-independent Vilkovisky-De Witt effective action
formalism [10] (see [11] for a general review) has been very useful
in
the investigation of gauge-independent spontaneous compactification
in
multidimensional gravity [12,13] based on the Einstein- or
 $R^2$-gravity action. It would be of interest to discuss the same
question for the still unknown $D\geq 4$ induced gravity (some
properties of 4D induced quantum
gravity  have been studied in refs. [14]). As a first step in this
direction, we here address the essential question: does the quantum
spontaneous compactification program actually work for 2D induced
gravity on an $R^1 \times S^1$ background?
Identifying $S^1$ with the time coordinate we discover another
motivation for the present study, namely 2D induced gravity at non-zero
temperature.
As we shall prove, the answer to the compactification question will be
positive.

The organization of the paper is as follows. In section 2 we review
the procedure of construction of
the gauge invariant effective action. This is particularized in
section 3 to the case of 2D induced gravity at one loop
order. In the gauge independent effective action, only a linear explicit
dependence on a constant parameter, $a$, remains. The actual
calculation of the traces involved in the expression for the action
is
carried out in section 4. A non-trivial minimum is found. Finally,
section 5 is devoted to conclusions.
\bs \bs

\section{The gauge-independent effective action}
\ms

In this section we construct the one-loop gauge-invariant effective
action for 2D induced gravity. Let us remember that the one-loop
conventional (i.e., gauge dependent) effective action is given by
\beq
\Gamma_{conv}^{(1)} (\phi ) = S(\phi ) +\frac{1}{2} \sp \ln S_{,ij}
(\phi ),
\eeq
where $S(\phi )$ is the classical action, $ \phi$ is the background
field, and euclidean notation is assumed everywhere. In the
background
field method it is not necessary for $\phi $ to be a solution of
the
equations of motion.

According to Vilkovisky and De Witt [10], the gauge-independent
effective action can be obtained by the method of replacing the
ordinary
functional derivative by the covariant functional derivative (in
one-loop approximation)
 \beq
S_{,ij} \longrightarrow S_{;ij} = S_{,ij} - \Gamma_{ij}^k S_{,k},
\eeq
where the condensed notation has been used and $\Gamma_{ij}^k$ is
the
connection in the space of fields. The term $S_{V,ij} =
-\Gamma_{ij}^k
S_{,k}$ is sometimes called the Vilkovisky correction.
It is very convenient to construct the connection using the metric
$\gamma_{ij}$ in the space of fields (configuration-space metric).
For
non-gauge theories, it can be constructed as the Christoffel
connection
[10]:
\beq
\Gamma_{jk}^i = \left\{_{jk}^i \right\}
= \frac{1}{2} \gamma^{il} \left( \gamma_{lj,k} + \gamma_{lk,j} -
\gamma_{jk,l} \right).
\eeq
The rule to define $\gamma_{ij}$ has been given by Vilkovisky [10].

In gauge theories, the construction of the connection is more
complicated. The physical field space is in gauge theories
different
from the naive field space, because of the local gauge symmetry.

Let $\gamma_{ij}$ be the metric of naive field space
\beq
ds^2 = \gamma_{ij} \delta \phi^i \delta \phi^j.
\eeq
By projecting $\delta \phi^i$ onto the physical field space, we get
\beq
\delta \phi_{\perp}^i \equiv \Pi_{j}^i \delta \phi^j, \ \ \ \ \Pi_{
j}^i \equiv \delta_j^i - R_{\alpha}^i N^{\alpha \beta} R_{\beta}^k
\gamma_{kj},
\eeq
where $R_{\alpha}^i$ is the generator of the gauge symmetry (i.e.,
$\delta \phi^i = R_{\alpha}^i \epsilon^{\alpha}$), and we have
$\Pi_j^i
\Pi_k^j = \Pi_k^i$, $\Pi_j^i R_{\alpha}^j =0$, and $N_{\alpha
\beta} =
\gamma_{ij} R_{\alpha}^i R_{\beta}^j$, $N^{\alpha \beta}$ being the
inverse of $N_{\alpha \beta}$.

Taking this into account, the metric on the physical field space is
\beq
ds_{\perp}^2 = \gamma_{ij} \delta \phi_{\perp}^i \delta
\phi_{\perp}^j = \gamma_{ik} \Pi_j^k \delta \phi^i \delta \phi^j .
\eeq
Using the new metric $\gamma_{ik} \Pi_j^k$, the connection
$\Gamma_{ij}^k$ for the physical field space is then given by
[10,11]
\beq
\Gamma_{ij}^k = \left\{_{ij}^k \right\} + T_{ij}^k, \label{7}
\eeq
where $ T_{ij}^k = -2B_{(i}^{\alpha} D_{j)} R^k_{\alpha} +
R_{\sigma}^m D_m R^k_{\tau} B_{(i}^{\sigma} B_{j)}^{\tau} $,
$B_i^{\alpha} =\gamma_{ij} N^{\alpha \beta} R_{\beta}^j$, $
D_jR^k_{\alpha} \equiv R^k_{\alpha ;j}$.

Finally, for gauge theories the one-loop gauge-independent
effective action is given by [10,11]
\bea
\Gamma^{(1)} &=& S(\phi ) +\frac{1}{2} \sp \ln \left[ S_{,ij} (\phi
) + S_{GF,ij} (\phi ) \right. \nn \\
&-& \left. \Gamma_{ij}^k(\phi ) S_{,k} (\phi ) \right] - \sp \ln \left[
R_{\alpha}^i (\phi ) \chi^{\alpha}_{,i} (\phi )\right], \label{8}
\eea
where $\chi^{\alpha}$ is the linear gauge condition, $S_{GF} =
\frac{1}{2} \beta_{\mu \nu} \chi^{\mu} \chi^{\nu}$, and $\beta_{\mu
\nu}$ is background-field independent (for more details see
[10,11]). As it can be checked explicitly, the one-loop effective
action (\ref{8}) is parametrization invariant, gauge invariant, and
gauge-fixing independent.
\bs \bs

\section{One-loop action for induced 2D gravity}
\ms

Let us now consider  induced 2D gravity, with the action
\beq
S= \int d^2x \, \sqrt{g} \, \left( R \frac{1}{\Delta} R + \Lambda
\right), \label{9}
\eeq
on the background $R^1 \times S^1$.
On such a background, which is not the solution of the classical
equations of motion, the convenient effective action is always gauge
dependent. However, the $S$-matrix (the effective action {\sl on shell},
i.e., at the stationary points) is independent on the gauge condition
choice. Actually, we are working in the loop expansion, what leads to
explicit gauge dependence even on shell (perturbatively). This is why we
prefer to work with the gauge-independent effective action.

 In accordance with the standard
background field method, we split the metric as
\beq
g_{\mu \nu} \longrightarrow  g_{\mu \nu} + h_{\mu \nu},
\eeq
where $g_{\mu \nu}$ is the metric of flat space $R^1 \times S^1$
and $h_{\mu \nu}$ is the quantum gravitational field.

The gauge fixing action is chosen as
\beq
S_{GF}= \frac{1}{\alpha} \int d^2x \, \sqrt{g} \, \left(
\nabla_{\mu} h_{\rho}^{\mu} - \beta \nabla_{\rho} h \right)^2,
\eeq
where $\alpha$ and $\beta$ are the gauge parameters and $h =
h_{\mu}^{\mu}$.

Now, one can calculate the following terms of (\ref{8}) for the
present case
\bea
& & S_{,ij} (\phi ) + S_{GF,ij} (\phi )  \equiv  \frac{\delta^2 (S+
S_{GF})}{ \delta h_{\mu \nu} \delta h_{\rho\sigma}} \nn \\
& & = \frac{\nabla^{\mu} \nabla^{\nu} \nabla^{\rho}
\nabla^{\sigma}}{ \Delta} + 2 \left( \frac{\beta}{\alpha} -1
\right) \delta^{\rho \sigma}  \nabla^{\mu} \nabla^{\nu} + \left( 1-
\frac{\beta^2}{\alpha} \right) \delta^{\rho \sigma} \delta^{\mu
\nu} \Delta \nn \\
& & - \frac{\Lambda}{4}  \delta^{\mu \rho} \delta^{\nu \sigma}
 - \frac{1}{\alpha}  \delta^{\nu \sigma}  \nabla^{\mu}
\nabla^{\rho} + \frac{\Lambda}{8}  \delta^{\mu \nu}
\delta^{\rho\sigma}.  \label{12}
\eea
In the rhs of (\ref{12}), the symmetrization $(\rho \sigma)
\leftrightarrow (\mu \nu )$ is understood.

The problem  now  is to define the configuration-space metric for
the theory (\ref{9}). In accordance with ref. [10], the
configuration-space metric for quantum gravity is given by
\beq
\gamma_{ij} \equiv \gamma_{g_{\mu \alpha} g_{\nu \beta}} =
\frac{1}{2} \sqrt{g} \left( g^{\mu\alpha} g^{\nu\beta} +
g^{\mu\beta} g^{\nu\alpha}- ag^{\mu\nu} g^{\alpha\beta} \right),
\label{13}
\eeq
where $a$ is a constant parameter. The $a$-dependence of the gauge
independent effective action (so-called configuration-space metric
dependence) has been discussed in refs. [15]. It is interesting to
notice that 2D induced gravity is related with topological gravity [3].
The fact that the gauge invariant effective action in this theory
depends on the configuration-space metric probably means that the
off-shell
quantum field theory under investigation does actually depend on
this metric. Different choices for this configuration-space metric
can probably lead to  inequivalent quantum theories off shell
[3,16].

It is not difficult to find the connection $\Gamma_{jk}^i$
(\ref{7}) using the configuration-space metric (\ref{13}). On the
other hand, the term corresponding to the Vilkovisky correction is
here
\beq
S_V = - \Gamma_{g_{\mu\nu}(x) g_{\alpha\beta}(y)}^{
g_{\rho\sigma}(z)} h^{\mu\nu}(x) h^{\alpha\beta}(y) \, \frac{\delta
S}{\delta g^{\rho\sigma}(z)},
\eeq
and it can be written (according to (\ref{8})), as
\bea
\frac{\delta^2S_V}{\delta h_{\alpha\beta} h_{\rho\sigma}} &=&
\frac{\Lambda}{4} \left[ \frac{3-2a}{8(a-1)} g^{\alpha(\rho}
g^{\sigma) \beta} + \frac{-a^2+2a-2}{2(a-1)(2-a)} g^{\rho\sigma}
g^{\alpha \beta} \right. \nn \\
&-& \frac{1}{8(a-1)\Delta} \, \left( g^{\alpha (\rho}
\nabla^{\sigma )} \nabla^{\beta} +  g^{\beta (\rho} \nabla^{\sigma
)} \nabla^{\alpha}  \right) \\
&+& \left. \frac{a}{2(a-1)(2-a)} \left( g^{\rho\sigma}
\frac{\nabla^{\alpha} \nabla^{\beta}}{\Delta} + g^{\alpha\beta}
\frac{\nabla^{\rho} \nabla^{\sigma}}{\Delta}  \right) - \frac{1}{2-
a} \, \frac{\nabla^{\rho} \nabla^{\sigma}\nabla^{\alpha}
\nabla^{\beta}}{\Delta^2}  \right]. \nn \label{15}
\eea
The ghost operator in (\ref{8}) (the last term in that expression)
has now the following form
\beq
\delta_{\nu}^{\mu} \Delta + (1-2\beta ) \nabla^{\mu} \nabla_{\nu}.
\eeq

Finally, collecting (12) and (15), taking into account (16), and
expressing
$\sp \ln$ for the non-minimal operators in terms of $\sp \ln$ for the
minimal operators (see the method developed in refs. [17]), we get
the following result for the one-loop action
\beq
\Gamma^{(1)} = 2\pi R S \Lambda + \frac{1}{2} \left[ \sp \ln \left(
\Delta + \frac{\Lambda}{4(2-a)} \right) - 2 \, \sp \ln \Delta \right].
\label{17}
\eeq
Here, $2\pi R$ is the length of the compactified dimension, while
$S=\int dx$ is the ``volume" of the space $R^1$. As we see
explicitly,  in eq. (17) the dependence on the gauge
parameters $\alpha$ and $\beta$ has disappeared. However, an
explicit dependence on the parameter $a$ remains in this
gauge-independent action (17).
\bs \bs

\section{Calculation of the traces and non-trivial minimum}
\ms

The trace calculations involved in expression (17) for the
one-loop effective action are somehow involved. Non-trivial commutations
of series have to be carried out. Using the techniques which have been
developed in [18] for the derivation of zeta functions corresponding to
partial differential operators, e.g. (already specified to $R^1\times
S^1$)
 \bea
\zeta_{-\Delta + m^2} \left(\frac{s}{2} \right) &=& -S \int_0^{\infty}
\frac{dk}{\pi}
\sum_{n=-\infty}^{+\infty} \left[ k^2 + \left( \frac{2\pi n}{\beta}
 \right) + m^2 \right]^{-s/2} \nn \\
&=& -\frac{S}{\sqrt{\pi}} m^{1-s} \left\{ \frac{-\Gamma
\left(\frac{s-1}{2}\right)}{2\Gamma \left( \frac{s}{2} \right)} +
\frac{\beta m}{2\sqrt{\pi}} \frac{1}{s-2}
+  \frac{\left( \frac{\beta m}{2} \right)^{(s-1)/2}}{
\Gamma \left( \frac{s}{2} \right)}  \sum_{k=0}^{\infty} \frac{(16
\pi)^{-k}}{k!} \right. \nn \\
 &\times & \left. \left( \frac{2 \pi}{\beta m} \right)^k \,
 \prod_{j=1}^k \left[ (s-2)^2-(2j-1)^2 \right] \,
\sum_{n=1}^{\infty} n^{\frac{s-3}{2}-k} \, e^{-\beta mn} \right\},
\eea
we get
\bea
V&=& \frac{\Gamma^{(1)}}{S} = 2\pi R \Lambda + \frac{R\Lambda}{32 (2-a)}
\left[ 1- \ln \left( \frac{\Lambda}{4(2-a)} \right) \right] -
\frac{1}{8} \sqrt{\frac{\Lambda}{2-a}} + \frac{1}{24R} \nn \\
&-& \frac{1}{4\pi \sqrt{2R}} \left( \frac{\Lambda}{2-a} \right)^{1/4}
\sum_{k=0}^{\infty} \frac{ (16 \pi )^{-k}}{k!} \left( \frac{R}{2}
\sqrt{\frac{\Lambda}{2-a}} \right)^{-k}  \\
& \times &  \prod_{j=1}^k \left[ 4-(2j-1)^2 \right] \,
\sum_{n=1}^{\infty} n^{-(k+3/2)} \, \exp \left( - \pi R
\sqrt{\frac{\Lambda}{2-a}} \, n \right). \nn
\eea

This expression can be very much simplified if we look for the basic
variables of the problem. They are
\beq
x \equiv \frac{\Lambda}{4(2-a)}, \ \ \ \ \ y\equiv R \sqrt{x} =
\frac{R}{2} \sqrt{\frac{\Lambda}{2-a}}.
\eeq
We have in terms of them
\beq
V= \sqrt{x} \, \left[ 8\pi (2-a) y + \frac{y}{8} (1- \ln x) -
\frac{1}{4} +\frac{1}{24 y} - F(y) \right],
\eeq
where the function $F(y)$ is given by
\beq
F(y)= \frac{1}{4 \pi} \, \sum_{k=0}^{\infty}
 \frac{ (16 \pi )^{-k}}{k!} y^{-(k+1/2)}
  \prod_{j=1}^k \left[ 4-(2j-1)^2 \right] \,
\sum_{n=1}^{\infty} n^{-(k+3/2)} \, e^{ - 2\pi n y}.
\eeq
It is now clear that all the dependence of the action on $R$, $\Lambda$
and $a$ comes through the specific combination given by variable $y$,
but for a global factor, $\sqrt{x}$, and for the first term, which is
just linear in $a$.

To proceed with the compactification program, we are ready to impose
(as is done in multidimensional gravity)
that
\beq
\left\{ \brr{l} V(R,\Lambda ,a)=0, \\ \frac{\txs\partial V(R, \Lambda ,
a)}{\txs \partial R} =0. \err \right.
\eeq
The explicit $a$ dependence can be readily eliminated, and we get
\beq
\sqrt{x} \left[ F(y)-yF'(y)-\frac{1}{12y} + \frac{1}{4}\right] =0.
\eeq
This transcendent equation involves an asymptotic series, and must be
solved aproximately. Fortunately, the decreasing exponentials come to
rescue and, after an explicit calculation one gets the expected result:
\beq
y_1 = 0.33.
\eeq
This is the non-trivial stationary point of the effective action. The
trivial one is obtained for
\beq
x_0=0.
\eeq

As for the second derivative, we get
\beq
\frac{\partial^2 V}{\partial y^2} = \sqrt{x} \left[ \frac{1}{12y^3}
-F''(y) \right],
\eeq
where the explicit $a$-dependence has disappeared. Hence, this second
derivative has a definite sign (independent of $a$) at the stationary
point
\beq
\left. \frac{\partial^2 V}{\partial y^2}\right|_{y=y_1} \simeq 2 >0.
\eeq
Thus, the point is clearly a minimum, that we obtain for the following
combination of parameters:
\beq
\frac{\Lambda R^2}{2-a} \simeq \left( \frac{2}{3} \right)^2.
\eeq

\bs \bs

\section{Conclusions}
\ms

We have calculated the gauge-independent effective action in 2D quantum
gravity on the background $R^1\times S^1$. Considering this theory as a
toy model for $D\geq 4$ induced gravity, we have analyzed the
spontaneous
compactification conditions and have found that the minimum of the
effective potential is attained when the spontaneous compactification
conditions are fulfilled. We should note that this fact is a general
one. For example, if we use the convenient effective action the minimum
is  attained too (however, in that case the radius of spontaneous
compactification depends on the gauge parameters).

In multidimensional quantum gravity models on the background $R^{D-1}
\times S^1, \ D>2$, it has been found [12,13] that quantum spontaneous
compactification is unstable (a maximum of the effective potential is
reached). On the contrary, in 2D induced quantum gravity we have found
that spontaneous compactification is stable, on such a simple background
as $R^1\times S^1$. It would be of interest to understand the origin of
this good property of the theory. Maybe, one can guess that $D>2$
induced gravity (if it exists) should be realized as a Kaluza-Klein type
theory.

\vspace{2cm}

\ni{\large \bf Acknowledgments}

Discussions with T. Muta, H. Osborn and A.A. Tseytlin are greatly
appreciated. S.D.O. thanks the members of the Department E.C.M. of
Barcelona University for the kind hospitality.
This work has
been supported by Direcci"n General de Investigaci"n
Cient!fica y Tcnica (DGICYT), research projects
 PB90-0022 and SAB92-0072.

\end{document}